\documentclass[doublecol]{epl2}

\usepackage[latin1]{inputenc}
\usepackage{amsmath}
\usepackage{pstricks,pst-node,pst-circ,pst-plot}
\RequirePackage{dsfont}

\newcommand{\be}{\begin{equation}}
\newcommand{\ee}{\end{equation}}

\newcommand{\ii}{\ensuremath{\text{i}}}
\newcommand{\eu}{\ensuremath{\text{e}}}

\newcommand{\vek}[1]{\ensuremath{\boldsymbol{#1}}}
\newcommand{\einvek}[1]{\ensuremath{\widehat{\vek{#1}}}}

\newcommand{\beps}{\vek{\epsilon}} 
\newcommand{\bsigma}{\vek{\sigma}}
 
\newcommand{\btau}{\vek{\tau}}
\newcommand{\bd}{\vek{d}} 
\newcommand{\bk}{\vek{k}} 
\newcommand{\br}{\vek{r}}

\newcommand{\init}[1]{#1}
\newcommand{\rhoi}{\init{\varrho}}

\newcommand{\fin}[1]{#1'}
\newcommand{\rhof}{\fin{\varrho}}

\newcommand{\VV}{\ensuremath{\mathcal{V}}}

\newcommand{\mean}[1]{\ensuremath{\langle{#1}\rangle}}
\newcommand{\ketbra}[1]{\ensuremath{| #1 \rangle \langle #1 |}}
\newcommand{\ket}[1]{\ensuremath{|#1\rangle}}

\newcommand{\Eins}{\ensuremath{\mathds{1}}}
\DeclareMathOperator{\tr}{tr}

\newcommand{\protowit}{\ensuremath{M}}
\newcommand{\TAB}{T_{A,B}} 

\renewcommand{\theenumi}{\roman{enumi}}

\def\pattern{
  \begin{pspicture}(-3.9,-0.8)(4.75,2.8)
    \psset{xunit=.05,yunit=2.5,linewidth=1.5pt,dash=3pt 1.5pt,arrows=c-c}
    \psline{->}(0,0)(0,1)
    \psline{->}(-60,0)(85,0)
    \pnode(0,1){y}
    \pnode(85,0){x}
    \pnode(-65,.5){vis}
    \pnode(22.5,0){alpha}
    \psline[arrows=-,linewidth=1pt,linestyle=dotted](-60,.5)(85,.5)
    \nput[labelsep=.25]{0}{y}{$\mean{P}^{'}\!/I_0$}
    \nput[labelsep=.25]{0}{x}{$\phi$}
    \psline[linewidth=.75pt](45,0)(45,0.875)
    \nput[labelsep=.15]{-90}{alpha}{$\underbrace{\hspace{2.25cm}}_{%
      \textstyle\vphantom{A}\alpha}$}
    \nput{180}{vis}{\rput{-90}{$\underbrace{\hspace{1.85cm}}_{%
      \textstyle\vphantom{A}\rput{90}{2\VV}}$}}
    \rput[tr](-1.5,.37){\rput[t]{-90}{$\underbrace{\hspace{.65cm}}_{%
      \textstyle\vphantom{A}\rput[r]{90}{
        \VV\cos(\alpha)\negthickspace\negthickspace}}$}}
    \rput[l](1,.5){\psframebox[fillstyle=solid,fillcolor=white,linestyle=none,framesep=1pt]{$1$}}
    \psplot[plotpoints=500]{-60}{85}{x 45 sub 3
      mul cos 0.75 mul 1 add 2 div}
  \end{pspicture}
}
\def\abring{
  {\psset{unit=.75}\begin{pspicture}(-3.05,-2.1)(3.05,2.1)%
    \pscustom{%
      \moveto(3,0.25)%
      \psarc(0,0){2}{7.180756}{172.819244}%
      \psline(-3,0.25)%
    }
    \pscustom{%
      \moveto(3,-0.25)%
      \psarcn(0,0){2}{-7.180756}{-172.819244}%
      \psline(-3,-0.25)%
    }
    \pscircle(0,0){1.5}
    \rput{0}(0,1.75){\rnode{imp1}{1}}
    \pscustom[linewidth=2pt]{%
      \psarcn(0,0){1.5}{105}{75}%
      \psarc(0,0){2}{75}{105}%
      \closepath%
    }
    \rput{0}(0,-1.75){\rnode{imp2}{2}}
    \pscustom[linewidth=2pt]{%
      \psarc(0,0){1.5}{-105}{-75}%
      \psarcn(0,0){2}{-75}{-105}%
      \closepath%
    }
    {\psset{linewidth=1.5pt,arrows=->}
      \psarcn(0,0){1.75}{170}{115}
      \rput{0}(-1.75,1.75){$A$}
      \psarcn(0,0){1.75}{65}{10}
      \rput{0}(1.75,1.75){$A$}
      \psset{linestyle=dotted}
      \psarc(0,0){1.75}{-170}{-115}
      \rput{0}(-1.75,-1.75){$B$}
      \psarc(0,0){1.75}{-65}{-10}
      \rput{0}(1.75,-1.75){$B$}
    }
    \pscircle[fillstyle=solid,fillcolor=black](0,0){0.05}
    \pscircle(0,0){0.2}
    \rput{0}(0.5,0){$\Phi$}
  \end{pspicture}}%
}
\def\youngpaths{
  \begin{pspicture}(-1.6,-1.6)(2.25,1.55)
    \rput{0}(0,0){\pnode{a1}}
    \pscircle[fillstyle=solid](a1){0.35}
    \rput(a1){$1$}
    \rput{0}(1.0,1){\pnode{a2}}
    \pscircle[fillstyle=solid](a2){0.35}
    \rput(a2){$2$}
    {
      \psset{nodesep=16pt,arrows=->,linewidth=1.5pt}
      \rput{0}(0,-2.0){\pnode{starta}}
      \rput{0}(1.0,-2.0){\pnode{startb}}
      \rput{0}(-2,0){\pnode{enda}}
      \rput{0}(-2,1){\pnode{endb}}
      \ncline{starta}{a1}
      \naput{$A$}
      \ncline{a1}{enda}
      \naput{$A$}
      \naput{$A$}
      \ncline[linestyle=dotted]{startb}{a2}
      \naput{$B$}
      \ncline[linestyle=dotted]{a2}{endb}
      \naput{$B$}
    }
  \end{pspicture}
}
\def\cbspaths{
  \begin{pspicture}(-0.6,-1.6)(3.25,1.55)
    \rput{0}(0,0){\pnode{a1}}
    \pscircle[fillstyle=solid](a1){0.35}
    \rput(a1){$1$}
    \rput{0}(2,1){\pnode{a2}}
    \pscircle[fillstyle=solid](a2){0.35}
    \rput(a2){$2$}
    {
      \psset{nodesep=16pt,offset=3pt,arrows=->,linewidth=1.5pt}
      \ncline{a1}{a2}
      \naput{$A$}
      \ncline[linestyle=dotted]{a2}{a1}
      \naput{$B$}
      \rput{0}(1,-2.0){\pnode{start}}
      \rput{0}(3,-1.0){\pnode{end}}
      \ncline{start}{a1}
      \naput{$A$}
      \ncline[linestyle=dotted]{a1}{start}
      \naput{$B$}
      \ncline{a2}{end}
      \naput{$A$}
      \ncline[linestyle=dotted]{end}{a2}
      \naput{$B$}
    }
  \end{pspicture}
}

\title{Entanglement Witnesses from Single-Particle Interference}
\shorttitle{Entanglement Witnesses from Single-Particle Interference} 

\author{Torsten Scholak\inst{1,3} \and Florian Mintert\inst{2,3} \and Cord A. M\"uller\inst{1}}
\shortauthor{T. Scholak \etal}

\institute{                    
  \inst{1} Physikalisches Institut, Universit\"at Bayreuth, %
95440 Bayreuth, Germany\\
  \inst{2} Max-Planck-Institut f\"ur Physik Komplexer Systeme, %
N\"othnitzerstra\ss{}e 38, 01187 Dresden, Germany\\
  \inst{3} Physikalisches Institut, Albert-Ludwigs-Universit\"at Freiburg, %
79104 Freiburg, Germany
}
\pacs{03.67.Mn}{Entanglement}
\pacs{73.23.-b}{Electronic transport in mesoscopic systems}
\pacs{42.25.Hz}{Interference}

\abstract{
We describe a general method of realizing entanglement witnesses in terms of
the interference pattern of a single quantum probe.
After outlining the principle, we discuss 
specific realizations both with electrons in mesoscopic Aharonov-Bohm
rings 
and with photons in standard Young's double-slit or 
coherent-backscattering interferometers.}

\begin{document}

\maketitle

How much information can be gained about separate parts of a composite
quantum system by scattering a single probe? 
It has been known since the founding days of quantum mechanics \cite{Schrodinger:1935gd} that
there are correlations between the subcomponents of quantum
systems that cannot be understood in terms of classical
probabilities. 
More recently, 
the quest for 
theoretical tools that characterize entangled states
\cite{Bell:1964qz,Peres:1996dd,Horodecki:1996zp} 
has been spured by the prospect of utilizing the underlying quantum
parallelism for a new era of information processing
\cite{Nielsen:2000lc}.
However, realizing these theoretical techniques in laboratory
experiments is still a challenge
\cite{Bourennane:2004fu,Leibfried:2005qf,Walborn:2006tw}, 
essentially because a multitude of observables
has to be measured.
Consequently, it is  desirable to characterize entangled states with a minimal set of measurements. 

In this Letter, we study the case 
of a probe particle that
gathers interferential information on the subsystems, which is then read
out by a suitable measurement only on the probe. 
This principle has been used in the past, 
such as in Bragg scattering of X-rays by crystals, since certain types of
correlations (such as positions on a lattice) can be inferred
from the interference pattern of a quantum probe. 
It is by no means obvious, however, whether a
single-particle interference pattern can distinguish subtle
quantum correlations in entangled states from classical correlations
in separable ones. 
It was shown in the context of mesoscopic solid-state systems 
that the visibility of conductance oscillations of double quantum 
dots can be sensitive to their entanglement \cite{Loss:2000ay}. 
Here we put this result in a broader context and 
show generally how the entanglement witness expectation
value of bipartite qubit systems can be read off the two-way 
interference fringes of a single quantum probe. 
In essence, this method exploits quantum parallelism: 
if the probe particle is brought into a superposition of two states, each of
which interacts with one of the subsystems, 
this single particle gains information about both subsystems
and its interference pattern can reveal the desired information about
entanglement. 

Consider a bipartite quantum system with the two subsystems labeled $1$ and $2$.
There are several techniques to determine whether their common state $\varrho_{12}$ is entangled or not \cite{Enk:2007zh}.
Bell inequalities for example are designed to distinguish quantum from
classical correlations as predicted by local realism
\cite{Bell:1964qz,Clauser:1969th}. 
If the correlations between suitably chosen observables exceed a given
threshold 
value, then the underlying quantum state is entangled \cite{Aspect:1982ng,Rowe:2001gf}.
However, a single inequality detects only rather few states,
and there is %
even a multitude of entangled states 
that cannot be detected by
any Bell test \cite{Hyllus:2005bx}. 

There are %
entanglement measures that detect every %
entangled state without %
state-dependent adjustment
\cite{Peres:1996dd,Horodecki:1996zp,Wootters:1998pi}. 
However, they can only be evaluated if 
the entire density operator $\varrho_{12}$ is known.
Since quantum state tomography \cite{Matsukevich:2008bq} 
requires a complete set of measurements,
it becomes less and less feasible 
with increasing dimension of the subsystems' Hilbert spaces.

State tomography can be avoided with methods of direct measurement---%
at the expense of multiple simultaneous state preparation 
\cite{Peres:1991dd,Ekert:2002gd,Alves:2003hb,Brun:2004lc,Bovino:2005mq,Mintert:2007jl}.
Here, some 
entanglement  %
measures can be rephrased or approximated as the
expectation value of collective observables on %
several identically
prepared quantum states
\cite{Horodecki:2003rp,Carteret:2005ri,Walborn:2006tw}. 
Both quantum state tomography and direct measurements are
applicable for every quantum state, %
but this   %
advantage of %
universality %
is gained by cutting back on experimental feasibility.

Entanglement witnesses \cite{Horodecki:1996zp}, by contrast, require neither 
complete state tomography nor 
multiple parallel 
state preparation. 
By definition, a witness is an observable with positive expectation
values $\init{\mean{W}}= \tr_{12}\{\rhoi_{12}W\} \ge 0$ on all
separable states, but negative expectation values on certain entangled
states. 
A negative measurement outcome thus implies with certainty that the target state was entangled.
A single witness (just like a single Bell inequality) cannot detect
entanglement universally, but (in contrast to Bell inequalities) 
to any target state one can find a witness that detects its
entanglement  
\cite{Horodecki:1996zp}. 

Still, a witness is a global observable on the combined target system,
and it %
requires {\em several} local measurements on the subsystems to 
reconstruct the witness %
\cite{Blaauboer:2005ye,Faoro:2007xw}. 
Our aim is now to evaluate %
the target witness expectation value 
$\init{\mean{W}}$ %
by measuring a suitable observable of a {\em single} probe
particle that is prepared in an initial state $\varrho_p$ and then interacts with the
target. The evolution %
in the combined target-probe system  %
from initial state $\rhoi$ to final state $\rhof$ is given by
$\rhof= T \rhoi T^\dagger$, where the operator $T$ depends 
on the Hamiltonian of the system under study.  
In terms of $T$ the expectation value of a probe observable $P$ in the final state reads
$\fin{\mean{P}} = \tr\{P\rhof\} = \init{\mean{T^\dagger P T}}$.
Typically, $P$ will be the projector onto a certain final probe 
state, e.g., a photon's direction and polarization.  
In this case the expectation values of the (semi-definite) positive
operator $T^\dagger P T$ 
cannot be negative, so that an entanglement witness seems nowhere in sight.

However, the probe particle can be in a superposition of states,
for instance propagating through either of the two slits of a Young 
experiment. In other words, we allow that the evolution from initial
to final state 
contains an interference between two exclusive alternatives labeled
$A$ and $B$. 
The phase difference $\phi_A - \phi_B = \phi$ between the
interferometric
path alternatives $A$ and $B$ should be under control, and the
evolution is given in terms of  
$T = e^{i\phi_A} T_A +  e^{i\phi_B} T_B$
with path-conditioned operators $\TAB$.

\begin{figure}[t!]
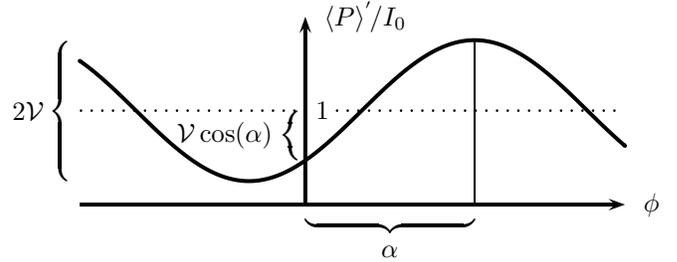
%
  \centering%
  \pattern%
  \caption{%
   Interference pattern of the probe expectation value. 
   For a suitable choice of probe parameters, destructive
   interference at the origin $\phi=0$ witnesses entanglement between
   the two target qubits.   
  }%
  \label{fig:pattern}%
\end{figure}

Spelling out all contributions, the probe expectation value on the
final state after interaction with the target becomes 
\be
\fin{\mean{P}} = \init{\mean{T_A^\dagger P T_A}} + \init{\mean{T_B^\dagger P T_B }}
+ \left[ e^{i\phi} \init{\mean{ T_B^\dagger P T_A }} + c.c. \right]. 
\label{eq:10}
\ee
The first two addends are phase-independent and sum up to the usual background intensity
$I_0$ 
of two-way interferometers.
In particular, for a positive probe observable $P$ 
these terms are also positive and by
themselves useless for constructing an entanglement witness. 
But there are also the interference terms, which are responsible for fringes as function of the phase
$\phi$ in the total detection intensity
\be
\label{pofphi}
\fin{\mean{P}} = I_0 \left[ 1 + \VV \cos(\phi-\alpha) \right]\ , 
\ee
drawn in
Fig.~\ref{fig:pattern}. 
The fringe visibility $\VV = (I_\text{max} - I_\text{min})/(I_\text{max} +
I_\text{min})$ and the interaction-induced phase shift $\alpha$ are determined by 
the expectation value of the first cross term in Eq.~(\ref{eq:10}), 
$2 \init{\mean{ T_B^\dagger P T_A }} = I_0  \VV e^{-i\alpha}$.
The real part of this quantity, i.e., 
the interference contribution 
at zero external phase shift $\phi=0$,
$I_0  \VV \cos \alpha$, 
is positive for constructive interference ($\cos\alpha>0$)  
and negative for destructive interference ($\cos\alpha<0$). It thus allows a sign
change that we can exploit in order to define a witness. 
Under the condition that probe and target states are prepared
independently, the initial state factorizes,
$\rhoi=\varrho_{12}\otimes\rhoi_{p}$, and 
the interference contribution at the origin  $\phi=0$
can be rewritten as 
$ \tr\{ \rhoi (T_B^\dagger PT_A + T_A^\dagger P T_B) \} =
\tr_{12}\{\rhoi_{12} M\}$, which is an expectation value of the target observable
\be\label{witness.def} 
  \protowit = \tr_{p}\left\{ \rhoi_{p} \left(T_B^\dagger  P T_A +
T_A^\dagger P T_B\right)\right\} \ .
\ee
This observable still depends on various quantities that can be chosen
in order to realize an entanglement witness: 
\begin{enumerate}
  \item the interference path alternative $A$, $B$,
  \item the path-conditioned interaction operators $\TAB$, 
  \item the initial probe state $\rhoi_{p}$, and 
  \item the probe observable $P$. 
\end{enumerate}
The mapping (\ref{witness.def}) from the simple probe system to an
observable of the composite target system 
constitutes the central idea of our Letter, together 
with the explicit examples, given below, for probe 
parameters such that $M$ is an entanglement witness.  
In essence, the initial state $\rhoi_{12}$ is entangled if the
interference pattern (\ref{pofphi}) of
the probe shows destructive interference at zero external phase
shift. 
Both visibility and phase are easily extracted by
fitting the experimental interference fringes. In this
respect, the two-way interference pattern of a single 
probe permits to extract subtle quantum correlations between
two given subsystems. 

In the following, we will consider the witnesses
\be
W_\pm = \Eins - 2 \, \ketbra{\Psi_\pm}\ ,
\label{eq:optWitWer}
\ee
in terms of the Bell states
$\ket{\Psi_\pm} = ( \ket{01} \pm \ket{10} ) / \sqrt{2}$,
which are the singlet and triplet states of total spin zero. 
The maximal overlap between a separable state and a Bell state is
$1/2$. Consequently, 
any separable state yields a positive expectation value of $W_\pm$
so that a negative expectation value reliably indicates entanglement.
In the remainder, we will show how these witnesses 
can be realized in mesoscopic solid-state devices by electron interference,
or for atomic systems by photon interference. 
It turns out that the singlet entanglement can be distinguished relatively
simply because of its distinct parity, but that it is much more difficult
to distinguish the entangled triplet state from its separable neighbors within
the triplet manifold.


\subsection{Solid-state realization}
 
\begin{figure}
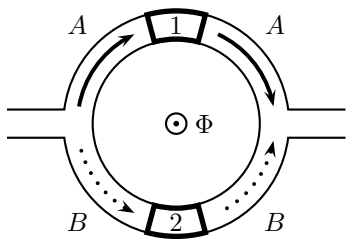

  \centering%
  \abring%
  \caption{%
    Electronic Aharonov-Bohm interferometer %
    with magnetic impurities embedded in each arm: %
    A mesoscopic ring threaded by a magnetic flux $\Phi$ %
    is attached to two supplying leads. %
  }%
  \label{fig:abring}%
\end{figure}

We shall first show that our concept allows to appreciate
a proposal for probing entanglement between resonant
quantum dots \cite{Loss:2000ay}, within a rather lucid model
description. 
Consider a single electron that probes 
the entanglement between two magnetic spin-1/2 impurities 
embedded in an Aharonov-Bohm (AB) ring
\cite{Washburn:1992ec,Pierre:2002yj},
depicted in Fig.~\ref{fig:abring}: (i) 
An electron propagating through arm $A$ interacts only with the first impurity,
whereas it interacts with the second impurity in arm $B$. 
The phase difference $\phi$ is controlled via the magnetic flux threading the ring, and the
interference fringes are recorded by measuring the conductance of the
ring. 

We model the effective interaction between the electron spin and the two impurities $j=1,2$
by the isotropic spin-flip Hamiltonian  
\be
\label{spinflip.eq}
V_j = \hbar g \bsigma \cdot \btau_j\ ,
\ee  
where $\bsigma=(\sigma^x,\sigma^y,\sigma^z)$ is the vector of Pauli matrices for the spin of the electron,
and $\btau_j$ are the impurity Pauli matrices.   
For this interaction type and geometry, the following choice of probe parameters (ii)--(iv) realizes the
entanglement witness $W_-$ of the singlet Bell state:  

(ii)
In a symmetric geometry the probe interacts with both impurities an
equal lapse of time $t$, the corresponding unitary time evolutions $T_A  =
\exp(-i gt \bsigma\cdot\btau_1)$ and 
$T_B  = \exp(-i gt \bsigma\cdot\btau_2)$, 
respectively, read
\be
  T_{A,B}   =\frac{\eu^{\ii gt}}{2} 
\left[ \left(\eu^{-\ii 2gt}+\cos 2gt\right )\Eins -\ii \sin(2gt)\bsigma \cdot \btau_{1,2}\right],
\ee
each with an interaction phase $gt$ that we take as an experimentally
tunable parameter. The partial probe trace of the cross
product $T_B^\dagger T_A$ is easily calculated to be 
\be
\tr_p\{T_B^\dagger T_A\} = \frac{1}{2}\left(|
\eu^{-\ii 2gt}+\cos 2gt|^2 \Eins + \sin^2(2gt)\btau_1\cdot\btau_2 \right)
\ .
\ee
The choice $2gt=\pi/2$ immediately yields the singlet witness of
(\ref{eq:optWitWer}) in the form 
$W_- = \frac{1}{2}(\Eins+\btau_1\cdot\btau_2)$.  
Consequently, with (iii) the probe spin in the 
unpolarized state $\rhoi_{p} = \frac{1}{2} \Eins$ and probe detection
without spin analysis, i.e., measuring  
(iv) the identity $P=\Eins$, the target operator (\ref{witness.def}) 
realizes the desired witness, $M=W_-$. 
Thus, we recover the prediction, derived in a more elaborate
theoretical description \cite{Loss:2000ay}, that AB current
oscillations across a singlet-entangled double quantum dot show a
characteristic minus sign. 

The triplet witness of (\ref{eq:optWitWer}) can only be written as an 
\emph{anisotropic} combination of Pauli matrices, 
$W_+ =
\frac{1}{2}(\Eins-\tau_1^x\tau_2^x-\tau_1^y\tau_2^y+\tau_1^z\tau_2^z)$.
Because 
both spatial symmetry of interaction and parity between the two
impurities have to be broken,     
the authors of \cite{Loss:2000ay} proposed to apply an inhomogeneous
magnetic field that provides the necessary (Berry) phase difference for the spin $x$-
and $y$-components along one of the two arms. 
In our description, one could equivalently resort to tuning the 
coupling strengths separately for each spin component in the spin-flip
interaction (\ref{spinflip.eq}). 
The triplet witness can then be realized by choosing the same 
coupling phase $2gt=\pi/2$ as before for all spin components except 
the $x$- and $y$-components of only one of the impurities which
should see a stronger coupling $2g't=3\pi/2$ or the reversed sign
$2g't=-\pi/2$. 

But instead of requiring finetuned coupling strengths or supplementary
control fields, we rather wish to realize the witness by measuring 
solely the probe particle. 
We have found that an effective witness for the triplet Bell state can be
realized using an initially polarized electron state 
$\varrho_p=\frac{1}{2}(\Eins + \sigma^z)$ and
measurement of the observable $P=\sigma^z$ such that 
$M = W_+ + \frac{1}{2} (\tau_1^z+\tau_2^z)$.
Its expectation value in the Bell state $\ket{\Psi_+}$ is $-1$,
whereas  
the expectation value in any separable state cannot be smaller than
$-\frac{1}{4}$, which sets only a slightly lower threshold value for entanglement
detection than zero. 
Furthermore, 
effective witnesses for the two other triplet Bell states 
$\ket{\Phi_\pm} = ( \ket{00} \pm \ket{11} ) / \sqrt{2}$ are
immediately obtained 
by a unitary rotation on $\varrho_p$ and $P$, 
from $\sigma^z$ to $\sigma^x$ or $\sigma^y$, respectively.

\subsection{Quantum optics realization}

In a second approach we examine interference of low-intensity laser light,
i.e.,\ a single probe photon scattered by two tightly
trapped atoms
on their resonant dipole
transitions of total angular momentum $\frac{1}{2}$
in ground and excited state.
In the absence of an external magnetic field, each degenerate ground
state is an effective spin-$\frac{1}{2}$, such that the two atoms carry a qubit pair. 
This situation corresponds to the experiments
by Eichmann et al.\ \cite{Itano:1998yi},
who studied 
how the Young-fringe visibility disappeared when path knowledge
was encoded in the atomic ground states.
The internal states of the atoms,
however, were separable, and the possible influence of their
entanglement was not studied.
 
The scattering of a photon with wave vector $\bk$ and transverse
polarization $\beps{\perp}\bk$ 
by a resonant atomic dipole transition is
described, within the dipole coupling scheme, by the dyadic operator
$\bd\circ\bd$. Acting on a pure spin-1/2 multiplet, the dipole vector operator $\bd$ 
must be proportional to
the only available vector operator, the spin operator itself. Thus we
can take $\bd=\btau$, 
up to a frequency-dependent prefactor describing the interaction
strength left implicit  in the following. 
Photon scattering $(\bk\beps)\mapsto(\bk'\beps')$ 
by a single atom is therefore described by
$\bar{\beps}' \cdot T_{A,B} \cdot \beps =
(\bar{\beps}'\cdot\btau_{1,2})(\btau_{1,2}\cdot\beps)$ where
$\bar{\beps}'$ denotes the complex conjugate of the scattered
polarization vector, and the scattering phase for an atom at the position
$\br_j$ reads $\phi_j=(\bk-\bk')\cdot\br_j$, $j=1,2$.

\begin{figure}
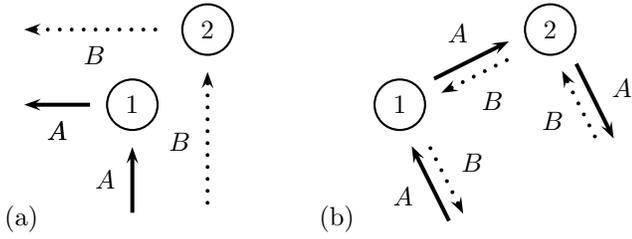

  (a)%
  \hspace{-1em}%
  \parbox[b]{0.45\columnwidth}{\youngpaths}
  (b)%
  \parbox[b]{0.45\columnwidth}{\cbspaths} 
  \caption{%
    Path alternatives for (a) Young interference and (b) coherent
    backscattering of a photon by two tightly trapped atoms.  %
  }%
  \label{fig:photonpaths}%
\end{figure}%

(i) We consider first the 
Young-type interference of Fig.~3(a), in which the %
scattering %
atoms constitute
the path alternatives. The phase
difference is given by $\phi = (\bk-\bk')\cdot(\br_1-\br_2)$.

(ii) 
Proceeding as in the solid-state realization above, we calculate the
interference contribution for an unpolarized initial photon state
$\rhoi_{p} = \frac{1}{2} \Eins$ and no polarization analysis at
detection $P=\Eins$, 
\be 
\tr_p\{T_B^\dagger T_A\} = 
\sum_{\beps'\perp\bk'}\sum_{\beps\perp\bk}
(\bar{\beps}\cdot\btau_2)(\btau_2\cdot\beps')
(\bar{\beps}'\cdot\btau_1)(\btau_1\cdot\beps)
\ .
  \ee
The sums over polarization vectors yield tranverse
projectors, $\sum_{\beps\perp\bk} \bar{\beps} \circ \beps = \boldsymbol{1}-\hat{\bk}\circ \hat{\bk}$, and the target observable (\ref{witness.def}) becomes 
\be
M = \bigl(1 + (\hat{\bk}\cdot\hat{\bk}{}')^2\bigr) \Eins 
+ \btau_1\cdot \boldsymbol{M} \cdot \btau_2 
\ee
with a dyadic $ 
\boldsymbol{M}= \hat{\bk}\circ \hat{\bk} + \hat{\bk}{}'\circ \hat{\bk}{}' +
(\hat{\bk}\times\hat{\bk}{}')\circ  (\hat{\bk}\times\hat{\bk}{}')$. 
For detection perpendicular to incidence,
$\hat{\bk}\cdot\hat{\bk}{}'=0$, $\boldsymbol{M}$ becomes the sum of projectors onto the three
orthogonal directions $\hat{\bk}$, $\hat{\bk}{}'$, and $\hat{\bk}\times\hat{\bk}{}'$
such that
$\boldsymbol{M}=\boldsymbol{1}$ is the \mbox{$3\times3$} unit matrix.  
Consequently, the Young
interference of (iii) an  unpolarized photon detected around right
angles from the incident direction and (iv) without polarization analysis
realizes the singlet witness up to an irrelevant 
multiplicative factor: $M = \Eins+\btau_1\cdot\btau_2 =2
W_- $. 

In this example, detection perpendicular to incidence is
necessary because all three spin components appearing in $W_-$ can only be probed
with tranverse photons if two directions of propagation are used.
 However, even for arbitrary directions of probe propagation and
arbitrary polarization states at incidence and detection, the Young
interference does not permit to realize the triplet witness.  

In order to realize $W_+$ by a projective measurement, we have to go beyond the Young interference
term, which is the leading single-scattering contribution in a general
multiple-scattering expansion. 
To next order in the small parameter $1/k|\br_1-\br_2|$,
we consider double scattering where the atoms exchange a single,
intermediate virtual photon. There are now again two path
alternatives, 
corresponding to the order of scattering events, see
Fig.\ \ref{fig:photonpaths}(b). Characteristically, the phase difference
$\phi= (\bk+\bk')\cdot(\br_1-\br_2)$ 
vanishes exactly for scattering in the backward direction $\bk'=-\bk$
such that the interference survives even an average over random
positions. This type of
interference explains both coherent
backscattering (CBS) in optics and weak
localization phenomena in mesoscopic electronic devices 
\cite{Akkermans:2007bf}.

Recently, the role of which-path information for CBS by atoms with
internal degeneracy \cite{Jonckheere:2000jy} has been studied
\cite{Miniatura:2007km}, 
but in the general case, no obvious signature of entanglement between
scatterers in the interference pattern was found. 
In the meantime, however, we could show that CBS can indeed be sensitive to
entanglement and allows to realize both witnesses $W_\pm$
with the following choice for the probe parameters: 

(i) For CBS interference, the path labels now describe the order in which the atoms are visited,
i.e.\ $1\to 2$ and $2\to 1$, respectively, as shown in Fig.~3(b).

(ii) 
The transition operator for path $A$, 
$  T_A =  \left( \btau_2 \circ \btau_2 \right)
    \cdot \left(  \boldsymbol{1} - \, \einvek{n} \circ \einvek{n} \right)
    \cdot \left( \btau_1 \circ \btau_1 \right)$, 
is obtained by connecting the 
single-scattering transition operators by the far-field projector
$ \left( \boldsymbol{1} - \, \einvek{n} \circ \einvek{n} \right) $
onto the plane transverse to the unit vector
$ \einvek{n} $ joining the two
atoms. 
The operator for path $B$ is obtained by exchanging the
role of the two atoms, i.e., by substituting 
$1 \leftrightarrow 2 $.

(iii)
The probe photon has to impinge
at right angles to the axis connecting the atoms
($\vek{k}\cdot\and\einvek{n}=0$) and again needs to be unpolarized. 

(iv)
The detection around the backscattering direction  $\bk'=-\bk$ can be made with a polarizing beam
splitter whose two outcomes realize both witnesses simultaneaously. 
The singlet witness $W_-$ can be found in the channel of linear
polarization along the unit vector
$ \einvek{n} $ joining the two
atoms, and the triplet witness $W_+$ in the channel of linear
polarization perpendicular to both $\einvek{n}$ and $\bk$. 

The CBS interference signal from two atoms can only be measured on top of the
single-scattering background, which contributes with a
geometry-dependent -- and generally much larger -- signal
strength and unfortunately the same fringe spacing as double
scattering. In the above configuration for the
$W_\pm$ witnesses, the single-scattering signal is non-zero, with a (Young) interference
visibility given by
$\VV=\frac{1}{2}(1+\mean{\tau_1^z\tau_2^z})$ and no phase shift
($\alpha=0$). But whereas single scattering contributes in
the completely mixed state with constructive interference ($\VV
=\frac{1}{2}$) completely masking the double-scattering
contribution, its visibility vanishes in the Bell states
$\ket{\Psi_\pm}$ ($\VV = 0$). Thus, 
if fringes with destructive phase are observed, they will be the CBS signal for the given entanglement witnesses on top of the flat
single-scattering background and therefore are the unambiguous signal
of bipartite entanglement. 

In summary, we have shown how the interference pattern of a single
probe that interacts with a two-qubit quantum system can 
witness its bipartite entanglement, the signature being the 
change from constructive to destructive
interference.
We have found proof-of-principle  models for realizations
in standard solid-state and quantum-optics settings, which now await
experimental realization and more quantitative calculations.

\acknowledgments
We thank Ch.~Miniatura
and B.-G.~Englert for pleasurable discussions.
F.M. acknowledges financial support by Alexander von Humboldt foundation.

\end{document}